\documentclass[%
 preprint,%
 amssymb, amsmath,%
 aip,cha,%
]{revtex4-1}

\usepackage{docs}%
\usepackage{bm}%
\usepackage[colorlinks=true,linkcolor=blue]{hyperref}%
\expandafter\ifx\csname package@font\endcsname\relax\else
 \expandafter\expandafter
 \expandafter\usepackage
 \expandafter\expandafter
 \expandafter{\csname package@font\endcsname}%
\fi
\usepackage{graphicx}
\graphicspath{{Images/}}
\usepackage[utf8]{inputenc}   
\usepackage{dcolumn}
\usepackage{color}
\usepackage{xcolor}
\usepackage{bm}
\usepackage{hyperref}

\newcommand{\change}[1]{\textcolor{blue}{#1}}	
\usepackage{amsmath}
\begin{document}


\title{Model for multi-shot all-thermal all-optical switching in ferromagnets}

\author{J. Gorchon}%
\email{jgorchon@lbl.gov}
\affiliation{Lawrence Berkeley National Laboratory, 1 Cyclotron Road, Berkeley, CA 94720, USA}
\affiliation{Department of Electrical Engineering and Computer Sciences, 
University of California, Berkeley, CA 94720, USA}

\author{Y. Yang}
\affiliation{Department of Materials Science and Engineering, University of California, Berkeley, CA 94720, USA}



\author{J. Bokor}
\affiliation{Lawrence Berkeley National Laboratory, 1 Cyclotron Road, Berkeley, CA 94720, USA}
\affiliation{Department of Electrical Engineering and Computer Sciences, 
University of California, Berkeley, CA 94720, USA}

\date{\today}

\begin{abstract}
All optical magnetic switching (AOS) is a recently observed rich and puzzling phenomenon that offers promising technological applications. However, fundamental understanding of the underlying mechanisms remains elusive. Here we present a new model for multi-shot helicity-dependent AOS in ferromagnetic materials based on a purely heat-driven mechanism in the presence of Magnetic Circular Dichroism (MCD). We predict that AOS should be possible with as little as 0.5\% of MCD, after a minimum number of laser shots. Finally, we reproduce previous AOS results by simulating the sweeping of a laser beam on an FePtC granular ferromagnetic film.
\end{abstract}

\keywords{All Optical Switching, Ferromagnets, Ultrafast, Magnetic Circular Dichroism}
\maketitle



The magnetic dynamics of a system triggered by an ultrashort optical pulse has been an exciting and yet unresolved problem in the magnetism community for the past 20 years. One of the most surprising results in the field was the discovery of the All-Optical Switching (AOS), the possibility of optically switching the magnetization of a magnetic material~\cite{Stanciu2007}. This was first observed in ferrimagnetic GdFeCo alloys~\cite{Stanciu2007}, later in a variety of ferrimagnet alloys and synthetic ferrimagnets~\cite{Mangin2014} and eventually in metallic ferromagnets~\cite{Lambert2014b}. It has been recently shown~\cite{Hadri2016} that two different types of switching need to be considered: First, a single-shot helicity-independent AOS found in ferrimagnets such as GdFeCo~\cite{Ostler2012}, and second, a multi-shot helicity-dependent AOS found in TbCo and ferromagnets~\cite{Hadri2016}. Although some models have been proposed that explain the switching in 2-sublattice systems such as ferrimagnets\cite{Radu2011,Ostler2012a,Barker2013,Atxitia2015}, modelling for AOS in ferromagnets is lacking.

In this letter, we present a model that describes multi-shot helicity-dependent AOS in ferromagnetic materials based on a purely heat driven mechanism. We first present the switching mechanism which is based on a combination of MCD and stochastic switching close to the Curie temperature $T_C$. This is followed by an in-depth description of the problem and of the way physical parameters are chosen. AOS is shown to be possible within a range of temperatures (i.e. laser fluences), for a large range of MCD values, but only after a certain number of laser shots. Finally, we reproduce previous AOS results by simulating the sweeping of the laser beam.

In our model, the mechanism driving the switching is a very simple and intuitive one: Whenever a laser heats a magnetic layer close to $T_C$, the stability of the magnetic state will be dramatically lowered as the anisotropy drops. For a circularly polarized beam, regions of the magnet with opposite magnetization will absorb different amounts of light due to MCD, resulting in hotter $T_{hot}$ and cooler regions $T_{cold}$. The difference in temperatures will lead to a difference in magnetic stability. If $T_{hot}\approx T_c$ and the MCD is large enough, cool regions will remain stable whereas hot regions will be prone to stochastic switching. Repeating the process (laser heating \& cooling) multiple times will statistically lead the magnet to full switching. 

\begin{figure}
\includegraphics[trim=0cm 0cm 0cm 0cm, clip=true, width=.60\columnwidth]{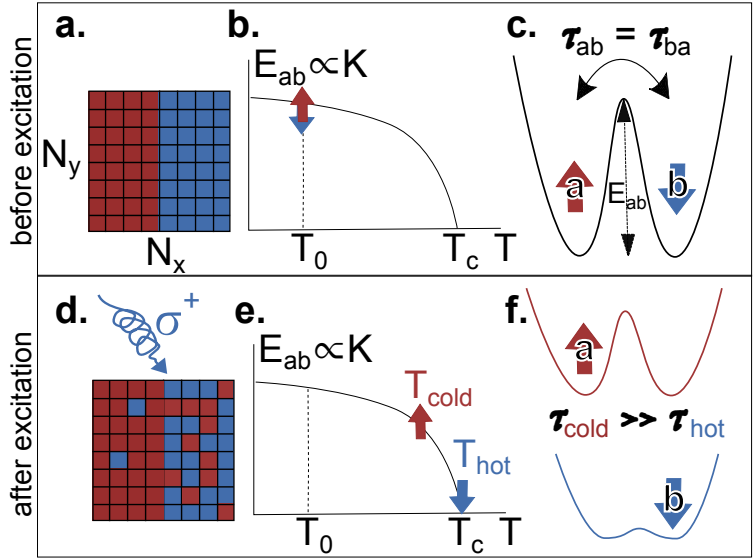} 
\caption{\label{fig:schema}: The AOS model supposes an a) $N_x$ by $N_y$ grid of cells, which present b) a temperature dependent magnetic anisotropy energy barriers $E_{ab}$, resulting in c) two stable possible states a (up) and b (down). As a d) circularly polarized laser pulse arrives onto the grid, e) cells in a and b states will absorb different amounts of energy due to MCD inducing a temperature distribution in the grid of hot and cold cells. This will lead to f) different energy landscapes for the hot and cold cells, resulting mostly in stochastic switching of the hotter cells.}
\end{figure}

In order to numerically test this idea, we represent the magnetic material by an array of $N_x$ by $N_y$ cells, grains or macrospins (as in Fig.~\ref{fig:schema}.a), presenting a strong out of plane magnetic anisotropy of energy density $K$. For the sake of simplicity we ignore exchange and dipolar couplings between cells. These hypotheses will be discussed later in the text. This means that the magnetic state of the cell can be represented by a symmetric double well potential, as shown in Fig.\ref{fig:schema}.c, where the magnetization can be only in the states \emph{up} or \emph{down}. The characteristic hopping time for the magnetization from state $a$ (\emph{up}) to state $b$ (\emph{down}) is given by the Néel-Brown formula~\cite{Brown1963}:
\begin{eqnarray}
\label{eq:tau}
\tau_{ab}(T) = \tau_0 e^{\frac{E_{ab}(T)}{k_B T}}
\end{eqnarray}
where $\tau_0$ is an attempt time typically estimated to be on the order of $~0.1$ ns~\cite{Lee2014}, $k_B$ is the Boltzmann constant, $T$ is the temperature of the cell and $E_{ab}$ is the energy barrier that prevents the magnetization from switching. The barrier will have the temperature dependence $E_{ab} (T) = K(T) V$ (depicted in Fig.~\ref{fig:schema}.b ) where $V$ is the volume of the cell. A high anisotropy at room temperature $T_0$ leads to long term stability of the magnetization. However, when heating the cells with a laser pulse (assuming a step-like heating profile of amplitude $\Delta T$ proportional to the laser fluence and duration $t_{hot}$) the probabilities that determine the final state ($a$ or $b$) of the magnetization when starting in state $a$ are given by\cite{KurtJacobs2010}: 
\begin{subequations}
\label{eq:Proba}
\begin{eqnarray}
P_{ab} = \frac{1}{2}(1-e^{\left({-\dfrac{t_{hot}}{\tau_{ab}(T_0+\Delta T)}} \right)})
\end{eqnarray}
\begin{eqnarray}
P_{aa} = 1 - P_{ab}
\end{eqnarray}
\end{subequations}
The probabilities are defined the same way when starting in state $b$, but the hopping time will be given by $\tau_{ba}$. The probability function spans from $0$ to $^1/_2$ as the energy barrier decreases from infinite to $0$. Intuitively, it means that when the cell is heated close to $T_C$ and the barriers disappear, the magnetization has no preferential direction. 

If the laser pulse is right $\sigma^+$ (left $\sigma^-$) circularly polarized~\cite{Comment2016}, $b$ ($a$) states will absorb more heat due to MCD. This difference will result in hot and cold cells where, assuming a temperature independent phonon heat capacity the temperatures are given by $T_{hot/cold}=T_0+(1\pm \frac{MCD}{2})\Delta T$ (Fig.~\ref{fig:schema}.e). Because of this difference in temperatures, hot and cold cells will have different switching probabilities (Fig.~\ref{fig:schema}.f) leading to a higher number of reversals of the hot cells (Fig.~\ref{fig:schema}.d). If we now heat the cells and allow them to cool back to $T_0$ $N$ times, the cumulative probability for the magnetization to end in state $b$ is given by (details in suppl. mat.~\cite{SuppMat}),

\begin{widetext}
\begin{eqnarray}
\label{eq:ProbaCum}
P_{B} &= \left(P_{ib}-\frac{P_{ab}}{P_{ab}+P_{ba}}\right) \left( 1 - P_{ab} - P_{ba}\right)^{N-1} + \frac{P_{ab}}{P_{ab}+P_{ba}}
\end{eqnarray}
\end{widetext}

where the subscript $i$ refers to the initial state. As $N$ increases, the probability is given by the last term in Eq.~\ref{eq:ProbaCum} and when $P_{ab} \gg P_{ba}$ we find $P_{B} \approx 1$. Consequently as long as enough heating cycles (i.e. laser pulses) are used and as long as there is a significant difference in the energy barriers for $a$ and $b$ cells, deterministic switching is expected. The barrier height difference originates in the difference in absorption due to MCD which leads to different temperature rises for cells in state $a$ and $b$. Because of this temperature difference and the strong~\cite{SuppMat} $dK/dT$ close to $T_C$, $E_{ab}$ will be very different for the two initial magnetization states.

Numerical simulations were conducted by considering an FePt-C-L1$_0$ granular film, for which AOS has been reported~\cite{Lambert2014b}. The typical size of grains is around $5$ nm wide and $7$ nm thick~\cite{Zhang2010}, where grains are separated by a $1$ nm thick C matrix~\cite{Rausch2015}. This matrix ensures thermal isolation, as well as magnetic exchange isolation. We can therefore safely neglect the exchange interaction and assume that important temperature distributions can exist. 

The temperature dependence of the magnetization $M_S$ and $K$ was extracted from Ref.~\cite{Thiele2002}, and corresponds to an Fe$_{50}$Pt$_{50}$ film. First $M_S$ was fitted with phenomenological equation~\cite{Fallis2013b} $ M_{s0}((T_c-T)/(T_c-T_0))^\gamma$ where $M_{s0}=1.15\cdot10^6$ A/m is the magnetization at $T_0=300K$, $T_c=775$ K and $\gamma=0.34$ is the phenomenological fitting exponent used for Fe~\cite{Fallis2013b}. Then $K$ was fitted with~\cite{Thiele2002,Staunton2004a} $ K_0(M_{S}/M_{S0})^2$ where $K_0=4.5\cdot10^6$ J/m$^3$. The fits are shown in the supplementary materials~\cite{SuppMat}.

The MCD was calculated, for a wavelength $\lambda=810$ nm, by using the non-magnetic complex index of refraction $n=3+4i$ and the complex non-diagonal term $\sigma_{xy}=-(1.4+1.7i)\cdot10^{14}$ s$^{-1}$ (c.g.s) of the optical conductivity tensor. These values were extracted from elipsometry, Kerr rotation and Kerr elipticity measurements in Refs.~\cite{Sato,Weller1997} through the relations reported in Refs.~\cite{Sato,XuThompson2002}. Through Maxwell equations, the complex index of refraction for left and right circular polarized light $n_\pm$ are found to be~\cite{XuThompson2002} $n_\pm=\sqrt{n^2 \pm 4\pi\sigma_{xy}/\omega}$, where $\omega=2\pi c/\lambda$ and $c$ is the speed of light. Fresnel equations were then used to obtain the reflectances and absorptions $A_+$ and $A_-$ for both helicities in the case of an infinitely thick film and normal incidence. Finally, the MCD was calculated as $2(A_+-A_-)/(A_++A_-)$. An MCD of $5.8$\% is obtained for Fe$_{50}$Pt$_{50}$.

\begin{figure}
\includegraphics[trim=0cm 0cm 0cm 2cm, clip=true, width=.55\columnwidth]{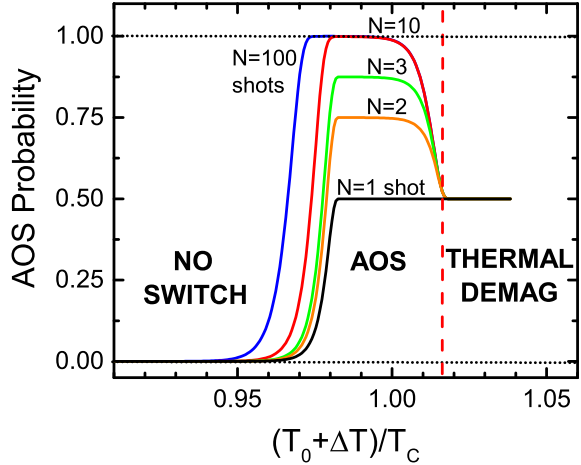} 
\caption{\label{fig:P_vs_T}: All-optical switching probability as a function of the laser temperature increase $\Delta T$ for different number of shots $N$. Three different temperature regimes are observed. At low temperature no switching is possible due to strong anisotropy barriers. Close to $T_C$ a certain amount of AOS occurs as the number of shots increases. With enough pulses full switching becomes possible. At higher temperatures the sample gets randomly demagnetized into a multidomain structure.}
\end{figure}

We first compute Eq.~\ref{eq:ProbaCum} as a function of the temperature, where the starting state $a$ is heated more than $b$ due to MCD. We begin with the calculated MCD$=5.8$ \%, a cell volume of $5$x$5$x$7=175$ nm$^3$~\cite{Zhang2010} and $t_{hot}=1$ ns. As shown in Fig.~\ref{fig:P_vs_T} for a single pulse ($N=1$, black line) the probability of switching is $0$ below a certain temperature threshold and $0.5$ above. No full switching is thus possible with a single shot. As we increase the number of shots, a narrow range of temperatures around $T_C$ results in a probability of switching that increases up to $1$ eventually ensuring full switching. In this case the absorbed critical fluences for AOS will be around $F= \Delta T h C\approx 1.16$ mJ/cm$^2$, where $\Delta T=T_C-T_0$, $h=7$ nm is the thickness and $C=3.5\cdot 10^6$ J/(m$^3$K) is the heat capacity of FePt~\cite{Kimling2014}.

Since this model assumes a sudden step-like temperature increase in the sample, cooling dynamics are not taken into account. Cooling in the presence of strong dipolar fields would make a full switching process less probable. However, this argument is consistent with the observation~\cite{Lambert2014b} that only thin films, with a small magnetization volume and thus smaller dipolar fields, exhibit nearly full AOS. Thicker films always show some degree of demagnetization (multidomains), and full switching is not observed.

In this calculation, the ratio $t_{hot}/\tau_0$ that acts as a prefactor in the exponential of Eq.~\ref{eq:Proba}.a was set equal to $10$. This parameter varies for different heat dissipation in the sample, but mostly offsets the temperature range at which AOS is observed (see suppl. mat.~\cite{SuppMat} for details).


\begin{figure}
\includegraphics[trim=0cm 0cm 0cm 0cm, clip=true, width=.55\columnwidth]{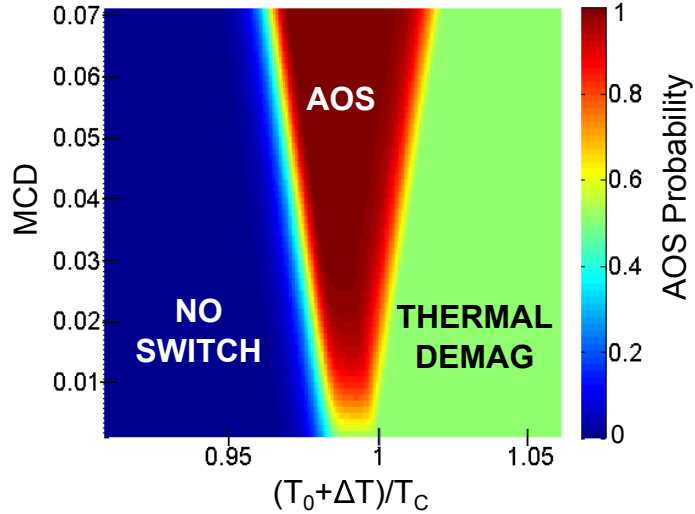} 
\caption{\label{fig:MCD_vs_T}: AOS Probability as a function of temperature and MCD for $N=10$ pulses.}
\end{figure}

As shown in Fig.~\ref{fig:MCD_vs_T}, the temperature window for AOS becomes larger as the MCD increases. However, even for a small MCD value of $0.5$\%, some AOS is still possible in a narrow range of temperatures.

\begin{figure}
\includegraphics[trim=0cm 0cm 0cm 0cm, clip=true, width=.55\columnwidth]{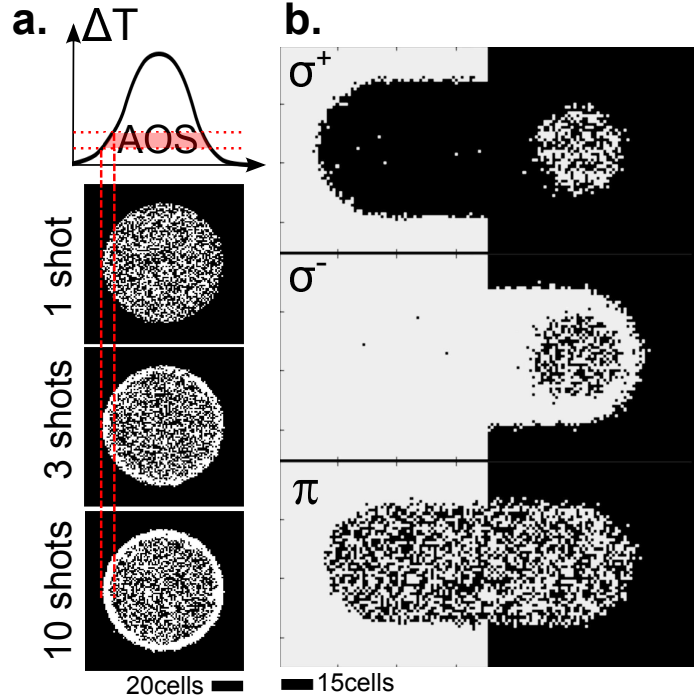} 
\caption{\label{fig:Sweep}: Simulations of the magnetization state after assuming a Gaussian temperature increase induced by the laser intensity profile (profile details in text). a) Simulation of successive shots with left circularly polarized ($\sigma^-$) light on an initially saturated black ("down") grid. After 10 shots a fully switched white ring develops. b) Simulations scanning the beam with left circular ($\sigma^-$), right circular ($\sigma^+$) and linearly polarized ($\pi$) laser shots at a \change{10} shots/cell sweeping speed. Initially, the grid consisted of a left-half \emph{up} (white) magnetization and right-half \emph{down} (black) magnetization. Each helicity favors one magnetization direction, which is determined by the edge of the Gaussian profile, whereas the linearly polarized laser beam only demagnetizes the sample (Multimedia view).}
\end{figure}

Next, the lateral spatial heating profile was assumed to be Gaussian, according to the laser intensity profile. The $1/e^2$ radius was set to $115$ cells and the temperature rise to $\Delta T=600$ (Fig.~\ref{fig:Sweep}.a) and $\Delta T=500$ (Fig.~\ref{fig:Sweep}.b).  As shown in Fig.~\ref{fig:Sweep}.a, with white and black corresponding to opposite magnetizations, one shot on an initially saturated grid results in a circular demagnetized pattern. As the number of shots increases an outer ring, within the temperature window for AOS, fully switches. This behavior resembles the AOS result in FePtC reported by Lambert el al.~\cite{Lambert2014b}.

We next scan the Gaussian beam (temperature profile) at a speed of 1 cell/$T$ where $T$ is the laser repetition period, and as shown in Fig.~\ref{fig:Sweep}.b (Multimedia view), we are able to write a magnetic domain by swiping the outer edge across the grid. Initially the grid was set so that the left half of the grid was \emph{up} (white) and the right one \emph{down} (black). Left circular ($\sigma^-$) and right circular ($\sigma^+$) polarized light causes \emph{up} and \emph{down} magnetizations respectively, whereas linearly polarized ($\pi$) light only demagnetizes the sample and results in multidomain states. Again, this result resembles previous results on AOS in FePtC by Lambert el al.~\cite{Lambert2014b}.

This AOS mechanism should thus operate in materials with large MCD, small dipolar fields, limited in-plane heat diffusion and a strong temperature dependence of the anisotropy close to $T_C$. We note that this doesn't restrict the mechanism to ferromagnets, thin ferrimagnetic films (such as TbCo, GdFeCo) or even antiferromagntic materials are also strong candidates for this switching mechanism due to their small dipolar fields and significant MCD. 

Finally we would like to discuss the validity of the model for ultrathin ferromagnetic films such as Pt/Co multilayers. These films do have significant MCD and exhibit low dipolar fields, however, they lack the granular structure that allows for magnetic and thermal isolation. Nevertheless, exchange interaction is not necessarily detrimental for the AOS. In such materials, we provide the following qualitative description: The first laser shot demagnetizes the sample. The magnetization spontaneously breaks into domains of various sizes. Under negligible dipolar fields, domain relaxation is dominated by the wall energy and pinning. The smallest domains will thus dissapear because shrinking forces induced by the wall energy increase as the bubble diameter decreases~\cite{Malozemoff1979}, whereas larger ones (100-1000 nm wide) will remain pinned. These larger domains will then remain cool on the next laser shot, while the rest of the film will restart the same process over. Repetition of this mechanism will eventually finish when various large domains merge together, resulting in full AOS.

In summary, we have proposed a new multi-shot all-thermal mechanism for helicity dependent AOS in magnetic materials, which is based on temperature distributions induced by the MCD. We calculated the cumulative probability for AOS after a certain number of pulses, and numerically estimated it for the case of an FePtC granular film. The AOS window as a function of MCD, temperature, and the number of pulses was presented showing that even with as little as $0.5$\% of MCD, multi-shot switching should still be possible. Finally our simulation of AOS using a scanned laser beam with different helicities being swept qualitatively reproduces previously reported experimental results.


See supplementary material for more details about the probabilities and chosen parameters.

This work was supported by the U.S. Department of Energy, Office of Science, Office of Basic Energy Sciences under Award Number DE-SC0012371 and the National Science Foundation Center for Energy Efficient Electronics Science.

\bibliography{library,library_MCD_AOS}

\begin{thebibliography}{24}
\bibitem [{Com()}]{Comment2016}%
  \BibitemOpen
  \href@noop {} {}\bibinfo {note} {Left, right circular and linear polarization
  are simulated by seting a positive, negative and null MCD,
  respectively.}\BibitemShut {Stop}%
\end{thebibliography}

\end{document}


\preprint{APS/123-QED}

SUPPLEMENTARY MATERIALS

\section{\label{sec:magnetic_parameters}AOS Probability after $N$ laser shots
}

Starting in state $a$ the probability of ending in state $a$ or $b$ is given by $P_{aa}$ and $P_{ab}$ respectively,

\begin{subequations}
\label{eq:Proba}
\begin{eqnarray}
P_{ab} = \frac{1}{2}\left(1-exp\left({-\dfrac{t_{hot}}{\tau_{ab}(T_0+\Delta T)}} \right) \right)
\end{eqnarray}
\begin{eqnarray}
P_{aa} = 1 - P_{ab}
\end{eqnarray}
\end{subequations}

where $\tau_{ab}(T)$ is the Néel-Brown dwell time (or hoping time),

\begin{eqnarray}
\label{eq:tau}
\tau_{ab}(T) = \tau_0 e^{\frac{E_{ab}(T)}{k_B T}}
\end{eqnarray}

The probability of ending in state $b$ after $N+1$ laser shots is equal to,

\begin{eqnarray}
\label{eq:PBNplus1}
P_B^{N+1} = P_{B}^{N}P_{bb} + P_A^NP_{ab}
\end{eqnarray}

Since we have $P_{bb}=1-P_{ba}$ and $P_A^N=1-P_B^N$, Eq.~\ref{eq:PBNplus1} becomes,

\begin{eqnarray}
\label{eq:PBNplus2}
P_B^{N+1} = P_B^{N}(1-P_{ab}-P_{ba}) + P_{ab}
\end{eqnarray}

This can be re-written as,

\begin{eqnarray}
\label{eq:PBNplus1X}
P_B^{N+1} + X = (P_B^N+X)(1-P_{ab}-P_{ba})
\end{eqnarray}

where X is equal to,

\begin{eqnarray}
\label{eq:X}
X = -\frac{P_{ab}}{P_{ab}+P_{ba}}
\end{eqnarray}

We can then write Eq.~\ref{eq:PBNplus1X} as,

\begin{eqnarray}
\label{eq:fracPB}
\frac{P_{B}^{N+1} + X}{P_{B}^{N}+X} = (1-P_{ab}-P_{ba})
\end{eqnarray}

which is equivalent to,

\begin{eqnarray}
\label{eq:fracPB2}
\frac{P_{B}^{N+1} + X}{P_{B}^{1}+X} = (1-P_{ab}-P_{ba})^N
\end{eqnarray}

After insertion of Eq.~\ref{eq:X} in Eq.~\ref{eq:fracPB2} and substituing $N$ by $N-1$ we obtain the total probability for ending in state $b$ after $N$ pulses,

\begin{eqnarray}
P_{B}^N &= \left(P_{B}^{1}-\frac{P_{ab}}{P_{ab}+P_{ba}}\right) \left( 1 - P_{ab} - P_{ba}\right)^{N-1} + \frac{P_{ab}}{P_{ab}+P_{ba}}
\end{eqnarray}

In the case where the initial state is $a$, the probability corresponds to a (All-Optical) switching probability.

\section{Magnetic properties of the FePt sample}

\begin{figure}
\includegraphics[width=.8\columnwidth]{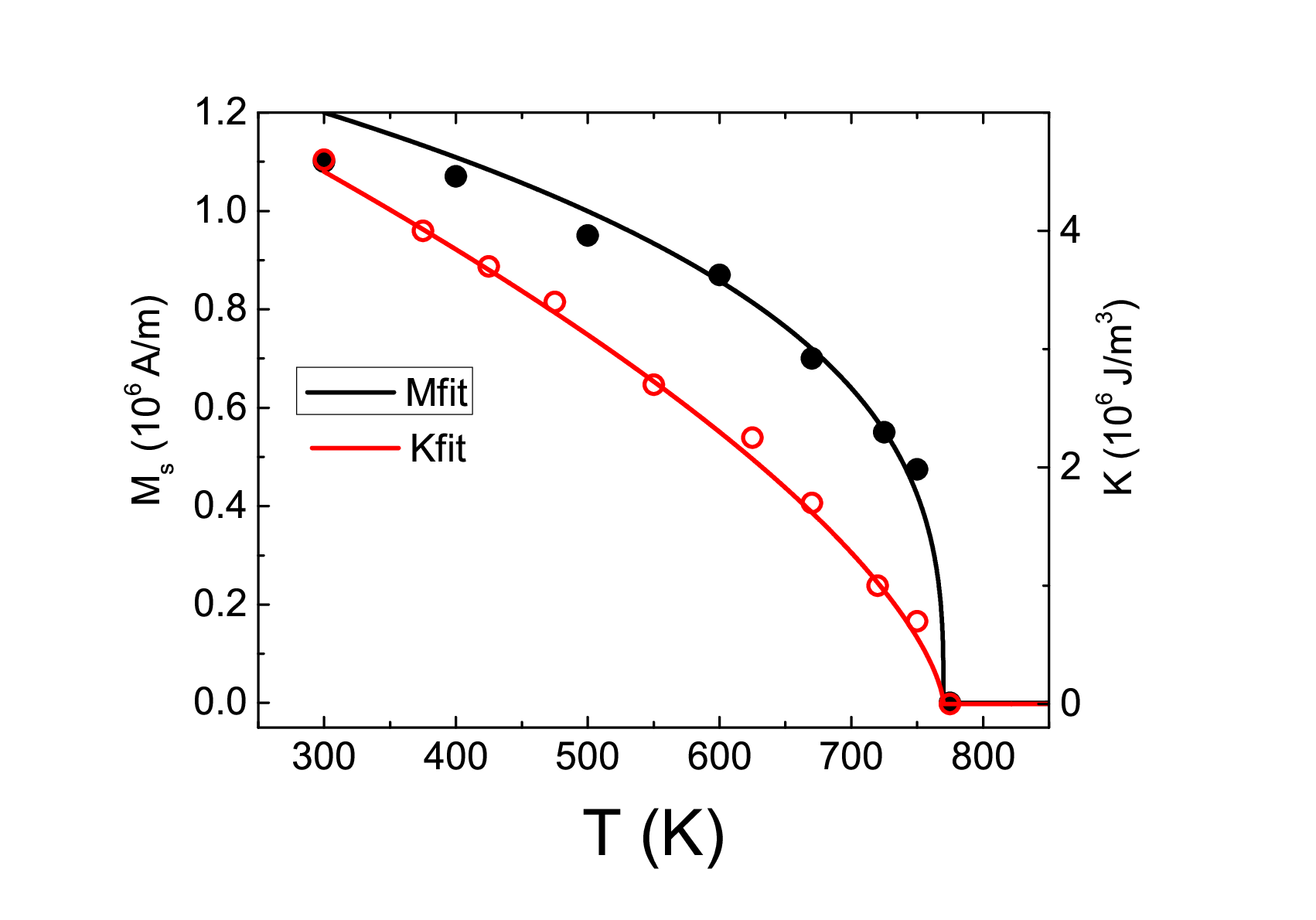}
\caption{\label{fig:Supp1}Magnetization $M_S$ and anisotropy $K$ as a function of temperature.}
\end{figure}

Fig.~\ref{fig:Supp1} shows the magnetization $M_S$ and anisotropy $K$ data versus temperature extracted from Ref.~\cite{Thiele} (dots) and the performed fits (solid lines). Fits are mostly valid~\cite{Fallis2013b} close to $T_C$. Fitting at lower temperatures is not relevant since the switching probabilities will only be affected close to $T_C$. The values from the fit to $K$ were then used for the simulations.

\section{Importance of $t_{hot}/\tau_0$}

\begin{figure}
\includegraphics[width=.8\columnwidth]{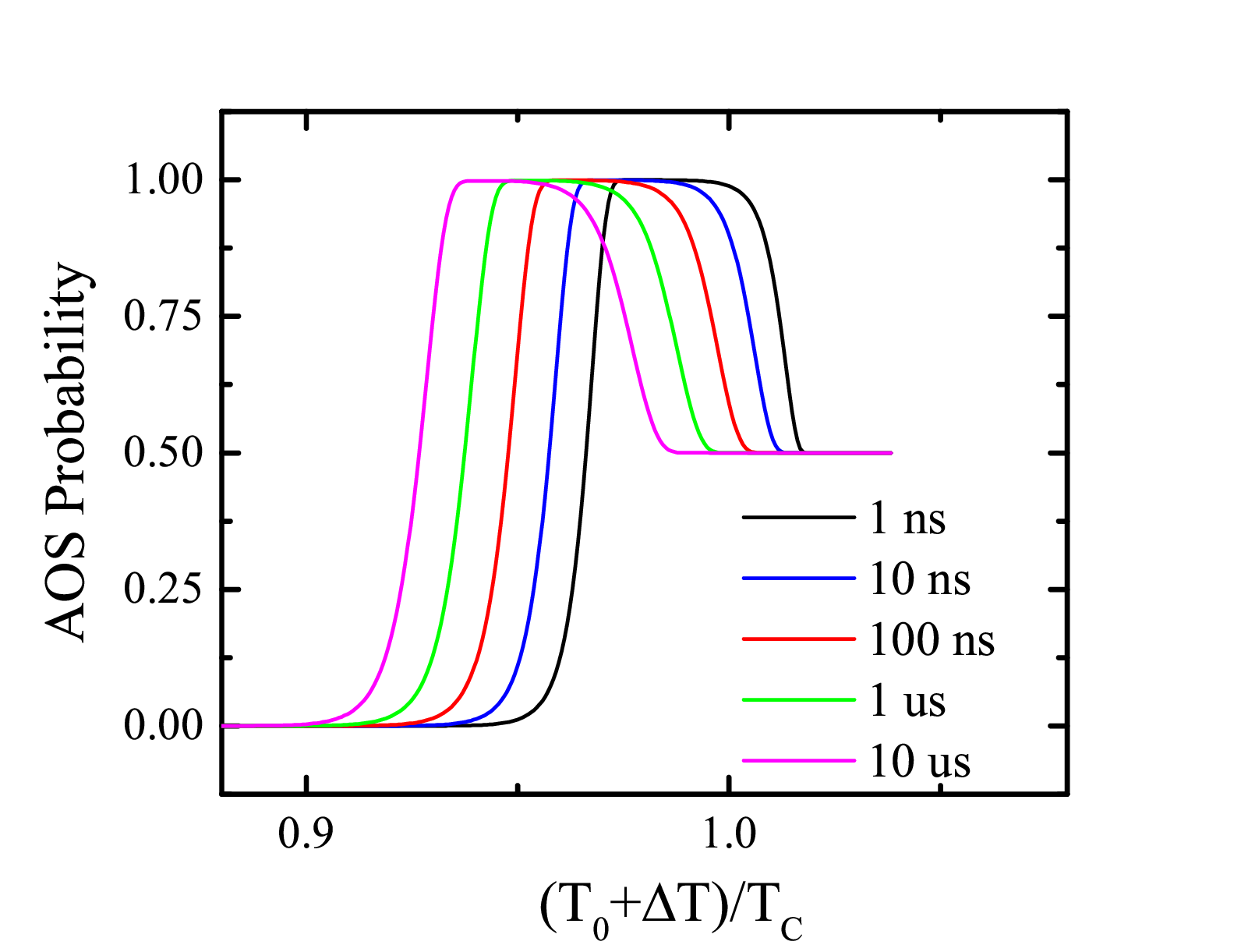}
\caption{\label{fig:Supp2} AOS probability after $N=100$ shots, as a function of the temperature rise $\Delta T$ for $t_{hot}$ ranging from $1$ ns to $10$ $\mu$s. $\tau_0$ is kept constant at $0.1$ ns.}
\end{figure}

As the time the system stays hot $t_{hot}$ increases (for a constant attempt time of $\tau_0=0.1$ ns), as shown in Fig.~\ref{fig:Supp2} the window where AOS is possible shifts towards smaller temperatures but keeps the same behavior. This is because as $t_{hot}$ increases the probability for switching will increase.

 
